\documentclass[12pt,paper,notoc,preprint]{JHEP}
\usepackage{epsfig}
\usepackage{axodraw}

\newcommand{\be}{\begin{equation}}
\newcommand{\ee}{\end{equation}}
\newcommand{\ba}{\begin{eqnarray}}
\newcommand{\ea}{\end{eqnarray}}
\newcommand{\dis}{\displaystyle}

\title{Scheme Dependence of Weak Matrix Elements in the $1/N_c$ 
Expansion} \author{Johan Bijnens\\Department of Theoretical
 Physics 2, Lund University,\\S\"olvegatan 14A, S 22362 Lund,
 Sweden}
\author{Joaquim Prades\\Departamento de F\'{\i}sica Te\'orica y 
del Cosmos, Universidad de Granada,\\Campus de Fuente Nueva, 
E-18002 Granada, Spain}
\abstract{We show how the scheme- and scale-dependence of the short-distance
operator product expansion with four-quark operators
can be correctly accounted for in the
framework of the $1/N_c$-expansion once the hadronization
of two-quark currents has been fixed. We show formulas explicitly 
in the case of the $B_K$-parameter. We then use them with our
earlier estimates of the long-distance effects. We compare Chiral Perturbation
Theory at Leading- and Next-to-Leading-Order with the ENJL model results
in the chiral limit. Good matching
between the long- and short-distance
 regimes is obtained and our final value for the physical
scheme independent $\hat B_K$ is $0.77\pm0.05\pm0.05$.}
\preprint{{LU TP 99--37}\\
{UG--FT--107/99}\\hep-ph/9911392\\November 1999}

\keywords{1/N Expansion, Kaon Physics, CP Violation}
\begin{document}
\section{Introduction}

Weak non-leptonic decays and mixings are one of our few  windows on
the CP-violating sector of the Standard Model and their calculation
is thus important. A set of reviews summarizing the present status
in the kaon area can be found in the various talks
at the workshops at Orsay and Chicago\cite{Orsay,Chicago}.

The large difference in mass between the $W$-boson and the kaon presents
an additional difficulty since logarithms of this ratio need to be
summed to all orders. This can be done using the Operator Product Expansion
(OPE) and is by now standard.
The remaining problem is the calculation of the matrix elements of these
operators at some low scale.

That the $1/N_c$-expansion would be useful in this regards was first suggested
by Bardeen, Buras, and G\'erard\cite{BBG87,BAR89} and has been reviewed
recently by Bardeen\cite{BAR99}.
There one can find most of the references to previous work
and applications of this non-perturbative technique.

Since the original
work\cite{BBG87} there are have been some improvements.
This has centered on identifying more correctly the
scales in short- and long-distance\cite{BBGpi,BGK} and their matching.
A more precise method to perform
this identification was proposed in \cite{BPBK},
the $X$-boson method,
and shown there to provide a correct matching with the renormalization
group evolution at one-loop. At next-to-leading (NLO) in the renormalization
group running several other problems appear. The operators become scheme
dependent and also dependent on various other choices like the one
of evanescent operators. This is discussed in the review by
Buras\cite{BUR98} and also in the original OPE at NLO
papers\cite{NLODeltaS=1,NLODeltaS=2}.

In this paper we  show how  scheme and scale dependence can
consistently be treated  within  the $1/N_c$-expansion technique
as argued in \cite{BAR89,BAR99,BP99,talks}.
We will use the method suggested in  \cite{BPBK} 
as  already used in  \cite{BP99,talks}.
In particular we will show explicit expressions
for the  current $\times$ current
$\Delta S=2$ four-quark operator in the
Standard Model. In practice this means tracking the scheme dependence
consistently during the entire calculation.

We are concerned here with the short-distance matching between 
the definition of the weak operators done in perturbative QCD 
using a particular regularization scheme and the actual calculation
of weak matrix elements at a given order in the $1/N_c$ expansion.
This can be done purely within perturbation theory and is
unambiguous once the short-distance schemes used in perturbative
QCD are specified. We will also show explicitly that it
does not depend on the infrared regulators.
We then use this result together with the matrix element obtained
earlier in \cite{BP99} to obtain complete scheme-independent results
for the $\hat B_K$ parameter within the $1/N_c$ expansion.
We also discuss the long-distance short-distance
matching for this quantity.

\section{The Flavour Changing Effective Action and the $X$-Boson Method}

In light hadrons flavour changing decays
 there appear two very different  scales;
namely, the hadronic scale and the weak scale, which 
makes necessary the use of effective field theory 
analyses.  This is done with the help of the OPE
  which allows   to separate
short-distance  from long-distance physics.
The OPE analysis is done at some scale {\em below} the
$W$ mass and this separation introduces some short-distance
scale and regulator scheme dependence which has to cancel 
in the final physical amplitude. 
We will see how this occurs
explicitly in the $1/N_c$ technique. 

The process of obtaining the relevant effective action
continues by integrating out the heavy
degrees of freedom up to some scale below
the charm quark mass and setting the appropriate
matching conditions at heavy particle
thresholds \cite{BUR98}.
This can be done using QCD perturbation theory.
In this  process, large logarithms
$\left(\alpha_s(\nu) \ln(M_W/\nu)\right)^k$ 
need to be resummed to all orders in perturbation
theory. This is done with the help of  Renormalization
Group (RG) techniques which at Next-to-Leading-Log-Order 
 (NLLO)  allow to resum up to $\alpha_s(\nu) \, 
\left(\alpha_s(\nu) \ln(M_W/\nu)\right)^k$.
In fact, only from NLLO can  a full scheme
dependence study be done. 

The whole process described schematically above
is by now standard and has been brought up to 
NLLO in \cite{NLODeltaS=1,NLODeltaS=2} from the earlier 
Leading-Log-Order (LLO) results\cite{LODeltaS=1}. 
For comprehensive reviews where complete
details can be found, see  \cite{BUR98}.

Thus, at some scale $\nu$ below the charm quark mass,
we are left with the effective field theory action
\be
\label{defGa}
\Gamma_{\Delta S=a} \equiv -C_{\Delta S=a}
{\dis \sum_{i}} \, C_i(\nu) \int {\rm d}^4 \, y \, \, Q_i(y)
+ {\rm h.c.}
\ee
to describe hadronic flavour changing processes in the Standard Model.
Here  $Q_i(x)$ is the set of four-quark operators made out
of the three light-quark  fields. They change flavour in $a$ units
and are defined perturbatively  within QCD using some regularization.  
In particular, the NLLO calculations  mentioned above have been done both
in the 't Hooft-Veltman (HV) scheme
($\overline{\rm MS}$ subtraction and 
 non-anti-commuting $\gamma_5$ in $D\neq 4$)
and in the Naive Dimensional  Regularization (NDR) scheme
($\overline{\rm MS}$ subtraction and anti-commuting $\gamma_5$ in $D\neq 4$).

In order to reach a physical process we now need to calculate matrix
elements of the effective action (\ref{defGa}) between the physical states.
In this calculation, all dependence on the  different schemes used
and on the scale $\nu$ should disappear. 
This necessarily involves long-distances which
cannot be treated analytically within QCD. One option is to simply
give up here and refer to lattice QCD  calculations. However,
these have at present many problems with this type of matrix elements.
They are complicated quantities to treat on the lattice
and chiral symmetry effects are very important. For a recent
review on lattice QCD weak hadronic matrix elements
see e.g. \cite{CFGLM99}. Here we would like to show how
to connect the effective action in a well defined way with other
approaches to low-energy QCD.

To identify a four-quark operator in any model or approximation to
QCD like e.g. Chiral Perturbation Theory (CHPT),  is very difficult.
 On the other hand, all these approximations and models are normally 
tuned to reproduce various experimentally observed properties involving 
currents and/or made to specifically match perturbative QCD predictions 
for Green functions involving currents. As a rule therefore, 
identification of two-quark currents is possible.
So a first step  is to replace $\Gamma_{\Delta S=a}$ by
an equivalent effective action that only involves currents.

We thus introduce an action of $X$-bosons with masses $M_X$
coupling to quark currents,
whose OPE reproduces Eq. (\ref{defGa}). This requirement fixes the couplings
of the $X$-bosons.
This matching can be done at a perturbative
level in QCD with
external states consisting of quarks and gluons
as long as the scale $\nu$ is high enough. At this matching the scale and
scheme-dependence of Eq. (\ref{defGa}) has disappeared and, as we show
below, the $X$-boson couplings are {\em independent} of the precise infrared
regulator chosen here.

We now split the integral over $X$-boson momenta into two parts.
The high energy part needs to have the momentum flow back through
perturbative quarks and gluons leaving an operator behind that only needs
to be evaluated to leading order in $1/N_c$. The low energy part needs to
be evaluated using the $1/N_c$ method to NLO in $1/N_c$.
Putting the two parts together we can see that all dependence on the $X$-boson
masses has dropped out and the final result can be written in a way that
only involves the coefficients of the operators in Eq. (\ref{defGa})
with the scheme dependence removed correctly.
In the next section, we will show this procedure explicitly for the
$\Delta S=2$ four-quark operator. The same procedure can be worked
out for the $\Delta S=1$ effective action but is much more
cumbersome\cite{BPP}.

\section{ The $\Delta S=2$ Case}

We  present  a complete analysis for the $\Delta S=2$ case. 
The generalization to other flavour changing processes is straightforward.
The Standard Model effective action for
$\Delta S=2$ transitions is 
\be
\Gamma_{\Delta S=2}\equiv - C_{\Delta S=2}
\, C(\nu) \, \int {\rm d}^4 \, y \, \, Q_{\Delta S=2}(y)
+ {\rm h.c.}
\ee
with 
\be
\label{QdeltaS=2}
Q_{\Delta S=2}(x) \equiv 4 L^\mu(x) L_\mu(x) \, ; 
{\hspace*{0.5cm}} 2 L^\mu \equiv
 \left[ \overline s \gamma^\mu 
\left(1-\gamma_5 \right) d \right](x) \, .
\ee
The operator $Q_{\Delta S=2}$
transforms under SU(3)$_L$ $\times$ SU(3)$_R$ rotations
as a 27$_L$ $\times$ 1$_R$. The
normalization factor 
\be
\label{defF}
C_{\Delta S=2}\equiv 
\frac{G_F}{4} {\cal F}(m_t^2,m_c^2,M_W^2, V_{CKM})
\ee
is a known function of the integrated out heavy particles masses
and  Cabibbo-Kobayashi-Maskawa matrix elements and
the Wilson coefficient $C(\nu)$ is known to NLLO\cite{NLODeltaS=2}.

The $\Delta S=2$ $K^0$-$\overline{K^0}$ matrix element  
\be
\label{KKB}
\langle \overline{K^0} | K^0 \rangle \equiv
- i \, C_{\Delta S=2} \, C(\nu) 
\langle \overline{K^0} | \int {\rm d}^4 \, y \, \, Q_{\Delta S=2}(y)
| K^0 \rangle \, 
\ee 
governs the short-distance contribution to the
$K_L$-$K_S$ mixing.  We  describe now how to calculate it
in the $1/N_c$ expansion \cite{BBG87,BPBK,BP99,Dortmund}.

\subsection{Transition to $X$-boson Effective Theory}
\label{Xbosontransition}

First of all, we need to set the
matching conditions between the weak operator $Q_{\Delta S=2}$
defined by (\ref{QdeltaS=2}) and the short-distance 
part of the effective field theory of $X$-bosons used in the
$1/N_c$ technique. The effective action of the latter is
\be
\label{defGX}
\Gamma_X \equiv  2 g_{\Delta S=2}(\mu_C, M_{X}, \cdots) \,
\int {\rm d}^4 y \, X^\mu(y) \, L_\mu (y) + 
{\rm h.c.}  \,\ee
where $g_{\Delta S=2}(\mu_C, M_{X}, \cdots)$ is an effective
coupling. In addition we add kinetic and mass terms for the $X$-bosons.
Evaluating matrix elements of $\Gamma_{\Delta S=2}$
needs to be done in the scheme in which it is defined.
We can freely choose the scheme in which we should treat $\Gamma_X$.
Here we choose an Euclidean cut-off regulator with subtraction to define
the strong coupling $\alpha_s$ at a scale $\mu_C$. This has certain advantages
for the next subsection.

We now determine the coupling $g_{\Delta S=2}$ 
by requiring that
the matrix elements of $\Gamma_{\Delta S=2}$ are the same as those
of $\Gamma_X$ to leading order in $1/M_X$.
We require at some perturbative scale
\be
\label{match}
\langle s(q_1) \overline d (q_4) 
| e^{i \Gamma_{\Delta S=2}} | \overline s (q_2) d(q_3)
 \rangle =
\langle s(q_1) \overline d (q_4) 
 | e^{i \Gamma_X}| \overline s (q_2) d(q_3)
 \rangle +{\cal O}(1/M_X^4) \;.
\ee
Where we have taken the matrix elements between an incoming strange 
anti-quark with momentum $q_2$ and an incoming down quark with 
momentum $q_3$ and an outgoing strange quark with momentum $q_1$ 
and an outgoing down anti-quark with momentum $q_4$.
We regulate infrared divergences by having $-q_i^2 > 0$.
In order to have expressions of manageable length we require
$-q_i^2\ll |q_i\cdot q_{j\ne i}| \ll \mu_C^2,\nu^2,M_X^2$ but we have kept all
momenta different in order to show the full cancellation and remove any
possible ambiguities in applying the equations of motion.

Both sides in (\ref{match}) are by themselves
scale, scheme, and regulator independent at at given order.
This is to order $\alpha_s^2(\nu)$ if we use NLLO results.
They do depend on the infrared regulators used,
like the quark momenta, and the gluon gauge
but these are the same
in both sides and cancel in the matching.
The gauge dependence we discuss in Section \ref{gauge}.

The left-hand side of (\ref{match})
evaluated at NLLO using the diagrams of Figure \ref{figOPEdiag}
\FIGURE{
\begin{picture}(90,90)(-45,-45)
\SetScale{0.5}                 
\ArrowLine(-80,80)(-4,4)
\ArrowLine(4,4)(80,80)
\ArrowLine(-4,-4)(-80,-80)
\ArrowLine(80,-80)(4,-4)
\GlueArc(-40,40)(28.28,-45,135){9}{3}
\Vertex(-60,60){2}
\Vertex(-20,20){2}
\CArc(0,0)(5.65,0,360)
\Line(-5.65,0)(5.65,0)
\Text(0,-45)[]{(a)}
\end{picture}
\hspace{-0.2cm}
\begin{picture}(90,90)(-45,-45)
\SetScale{0.5}                 
\ArrowLine(-80,80)(-4,4)
\ArrowLine(4,4)(80,80)
\ArrowLine(-4,-4)(-80,-80)
\ArrowLine(80,-80)(4,-4)
\Gluon(-60,60)(60,60){8}{5}
\Vertex(-60,60){2}
\Vertex(60,60){2}
\CArc(0,0)(5.65,0,360)
\Line(-5.65,0)(5.65,0)
\Text(0,-45)[]{(b)}
\end{picture}
\hspace{-0.2cm}
\begin{picture}(90,90)(-45,-45)
\SetScale{0.5}                 
\ArrowLine(-80,80)(-4,4)
\ArrowLine(4,4)(80,80)
\ArrowLine(-4,-4)(-80,-80)
\ArrowLine(80,-80)(4,-4)
\Gluon(-60,60)(-60,-60){-10}{5}
\Vertex(-60,60){2}
\Vertex(-60,-60){2}
\CArc(0,0)(5.65,0,360)
\Line(-5.65,0)(5.65,0)
\Text(0,-45)[]{(c)}
\end{picture}
\hspace{-0.2cm}
\begin{picture}(90,90)(-45,-45)
\SetScale{0.5}                 
\ArrowLine(-80,80)(-4,4)
\ArrowLine(4,4)(80,80)
\ArrowLine(-4,-4)(-80,-80)
\ArrowLine(80,-80)(4,-4)
\GlueArc(-60,-60)(120,0,90){8}{10}
\Vertex(-60,60){2}
\Vertex(60,-60){2}
\CArc(0,0)(5.65,0,360)
\Line(-5.65,0)(5.65,0)
\Text(0,-45)[]{(d)}
\end{picture}
\caption{The type of diagrams contributing to the matrix element
of $\Gamma_{\Delta S=2}$. The circle
denotes the current-current operator. Colours are connected on the same
side of the horizontal line. The curly line is a gluon and full lines are
quarks.
(a) Wave Function Renormalization.
(b) Vertex Diagrams. (c)(d) Box Diagrams.}
\label{figOPEdiag}
}
is
\be
\label{DIM}
i C_{D}\, 
\left[ \left( 1 + F(q_1,q_2,q_3,q_4)
\frac{\alpha_s(\nu)}{\pi}  \right) S_1 
+ \left( 1 + F(q_1,q_2,-q_4,-q_3)
\frac{\alpha_s(\nu)}{\pi}  \right) S_2  
\right] 
\ee
with 
\be
\label{defCD}
C_{D} =  - C_{\Delta S=2} \, C(\nu)
\left(  1 + \frac{\alpha_s(\nu)}{\pi} \left[ 
  \frac{\gamma_1}{2} \ln \left(\frac{2 q_1 . q_2}{\nu^2}\right)
+ r_1 \right] \right) 
\ee
where $\nu$ is a scale where perturbative QCD can be used.
The function
\be
F(q_1,q_2,q_3,q_4) =
B(q_1,q_2,q_3,q_4) - \frac{1}{N_c}
 B(q_1,q_2,-q_4,-q_3) + \frac{N_c^2-1}{2 N_c} 
V(q_1,q_2,q_3,q_4)  \, ; 
\ee
collects finite terms from the box (B)
and vertex diagrams (V).
\ba
S_1 &=& 
\left[\overline s_1 \gamma_\mu (1-\gamma_5) d_3 \right]
\left[\overline s_2 \gamma^\mu (1-\gamma_5) d_4 \right]
\nonumber \\
S_2 &=& 
 \left[\overline s_1 \gamma_\mu (1-\gamma_5) d_4 \right]
\left[\overline s_2 \gamma^\mu (1-\gamma_5) d_3 \right]
\ea
are  tree level matrix elements. The colour indices are summed
inside brackets.
 The $\Psi_i$ field destroys a quark $\Psi$  with momentum $q_i$.
In the NDR scheme, using the Feynman gauge
for the gluon, the one-loop anomalous dimensions is
\be
\gamma_1 =  \frac{3}{2} \left(1-\frac{1}{N_c}\right)
\ee
and
\be
\label{r1NDR}
r_1^{NDR} = -\frac{9}{4} \left(1-\frac{1}{N_c}\right) \, . 
\ee
The one-loop anomalous 
dimensions $\gamma_1$ is scheme-independent and
\be
\label{r1HV}
r_1^{HV} = -\frac{7}{4} \left(1-\frac{1}{N_c}\right) \, . 
\ee
if we use the 't Hooft-Veltman scheme instead.
In both cases we have used the same scheme for the evanescent operators
as the one in \cite{NLODeltaS=2,BW90}.

The terms with $F(q_1, q_2, q_3, q_4)$ collect the infrared
dependence on quark masses, external quark momenta,
$\cdots$ as well as scheme-independent constants.
If we had used a gluon mass as infrared regulator that would have been
present there as well.
The separation of the constant terms 
between $r_1$ and $F(q_1, q_2, q_3, q_4)$ is arbitrary. 

The explicit expression for the box
function $B(q_1, q_2, q_3, q_4)$ is
\ba
B(q_1, q_2, q_3, q_4) &=&
\frac{1}{4} \left[
\ln\left( \frac{-2 q_3 \cdot q_4}{q_3^2} \right)
\ln\left( \frac{-2 q_3 \cdot q_4}{q_4^2} \right) 
+ \ln\left( \frac{-2 q_1 \cdot q_2}{q_1^2} \right)
  \ln\left( \frac{-2 q_1 \cdot q_2}{q_2^2} \right)
 \right. \nonumber \\  &-&  
 \ln\left( \frac{-2 q_1 \cdot q_3}{q_1^2} \right)
 \ln\left( \frac{-2 q_1 \cdot q_3}{q_3^2} \right) 
- \ln\left( \frac{-2 q_2 \cdot q_4}{q_2^2} \right)
\ln\left( \frac{-2 q_2 \cdot q_4}{q_4^2} \right)
\nonumber \\ &+& \left. 
\frac{1}{2} \ln\left( \frac{q_1 \cdot q_3}{q_1 \cdot q_2} \right)
+ \frac{1}{2} \ln\left( \frac{q_2 \cdot q_4}{q_1 \cdot q_2} \right)
\right] 
+{\cal O}(q_i^2/(q_j\cdot q_{k\ne j}))
\ea
and for the vertex function  $V(q_1, q_2, q_3, q_4)$ is
\ba
V(q_1, q_2, q_3, q_4) &=&
-\frac{1}{4} 
\left[ 2+ \frac{4\pi^2}{3} +
 2 \ln\left( \frac{- 2 q_1 \cdot q_3}{q_1^2} \right)
  \ln\left( \frac{-2 q_1 \cdot q_3}{q_3^2} \right) 
\right. \nonumber \\  &+& \left. 
  2 \ln\left( \frac{-2 q_2 \cdot q_4}{q_2^2} \right)
\ln\left( \frac{-2 q_2 \cdot q_4}{q_4^2} \right) 
- \frac{3}{2}   \ln\left( \frac{-2 q_1\cdot q_3}{q_1^2} \right) 
-\frac{3}{2}  \ln\left( \frac{-2 q_1 \cdot q_3}{q_3^2} \right) 
\right. \nonumber \\  &-& \left. 
\frac{3}{2}  \ln\left( \frac{-2 q_2 \cdot q_4}{q_2^2} \right) 
-\frac{3}{2}  \ln\left( \frac{-2 q_2 \cdot q_4}{q_4^2} \right) 
\right] 
+ {\cal O}(q_i^2/(q_j \cdot q_{k\ne j}))\;.
\ea
In the HV scheme one gets the same expressions
for these two functions.

The Wilson coefficient $C(\nu)$ at NNLO
can be written as
\ba
C(\nu) &=&
\left(1+\frac{\alpha_s(\nu)}{\pi}\left[
\frac{\gamma_2}{\beta_1}-\frac{\beta_2 \gamma_1}{\beta_1^2}
\right]\right) \left[\alpha_s(\nu)\right]^{\gamma_1/\beta_1}
\ea
with \cite{NLODeltaS=2}
\ba
\label{running}
\gamma_2^{\rm NDR}&=& \frac{1}{32}
\left(1-\frac{1}{N_c}\right) \, \left[ -17 + 4 (n_f-3)+
\frac{57}{N_c}\left(1-\frac{N_c^2}{9}\right) \right]
\nonumber \\
\gamma_2^{\rm HV}&=& \frac{1}{32}
\left(1-\frac{1}{N_c}\right) \, \left[ -17 + 4 (n_f-3)+
\frac{57}{N_c}\left(1-\frac{N_c^2}{9}\right) 
- 16 \beta_1 \right]
\ea
with $\beta_1=-9/2$ and $\beta_2=-8$.
The Feynman gauge for the gluon has been used to obtain 
the results above \cite{NLODeltaS=2}.
The scheme of evanescent operators we used is the one of
\cite{NLODeltaS=2,BW90}.

Equation
 (\ref{DIM}) is now scheme-independent and scale-independent
to order $\alpha_s^2$, the difference in running via Eq. (\ref{running})
is compensated by the difference between (\ref{r1NDR}) and (\ref{r1HV}).
This was shown with simpler states also in \cite{BW90}.
We can now check this by plotting $-C_D/C_{\Delta S=2}$ versus $\nu$
for the two schemes. The difference and the variation with $\nu$
is an indication of the neglected  $\alpha_s^2$ corrections.
This is shown in Figure \ref{figCD}
where we plot the one-loop coefficient $C(\nu)$ and the two-loop
one for both the NDR and the HV schemes. They differ considerably
and are rather $\nu$-dependent. Now $-C_D/C_{\Delta S=2}$ of Eq. (\ref{defCD})
is more stable with $\nu$ and has a smaller scheme-dependence left.
Both are an indication of the size of the $\alpha_s^2$ corrections.
This can be improved systematically by doing the matching  to order
$\alpha_s^2$ and the running to three-loops and so on.
 We have chosen $2 q_1\cdot q_2 = 1$ GeV$^2$
and $\alpha_s(M_\tau)=0.334$ \cite{ALEPH} as input.
\FIGURE{
\epsfig{file=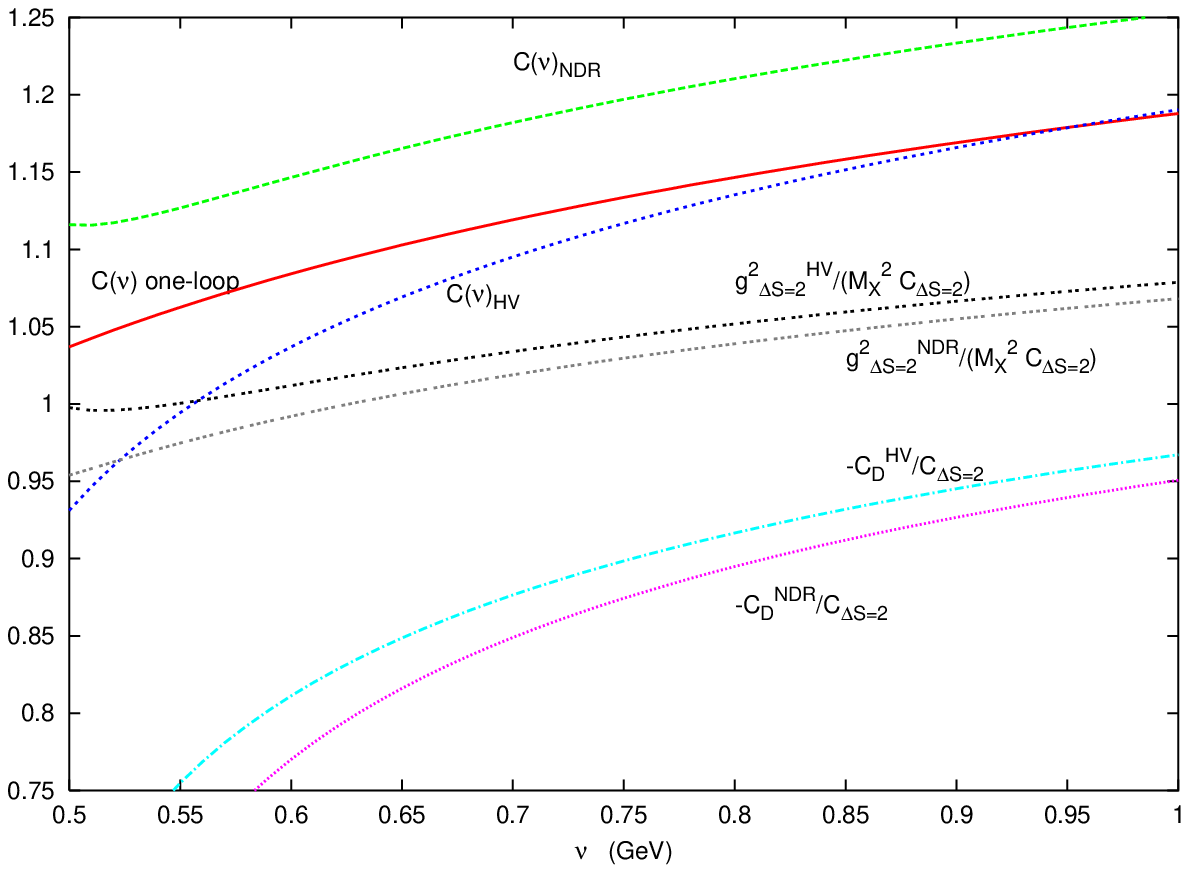,width=\textwidth}
\caption{The short-distance coefficients.}
\label{figCD}
}

For the left-hand side of (\ref{match})
we need to specify now  which is
the effective field theory we use in the $1/N_c$
calculation.   We introduced a fictitious
\cite{BAR89,BPBK} $X^\mu$ boson which reproduces
the physics of the $Q_{\Delta S=2}(x)$ weak operator 
{\em below} $\mu_C\approx\nu$, with coupling
 given in $\Gamma_X$ of Eq. (\ref{defGX}).

For the right-hand side of (\ref{match})
at NLLO we get
\FIGURE{
\begin{picture}(90,90)(-45,-45)
\SetScale{0.5}                 
\ArrowLine(-80,80)(0,40)
\ArrowLine(0,40)(80,80)
\ArrowLine(0,-40)(-80,-80)
\ArrowLine(80,-80)(0,-40)
\GlueArc(-40,60)(28.28,-23.5,156.5){9}{3}
\Vertex(-64,72){2}
\Vertex(-16,48){2}
\Vertex(0,40){2}
\Vertex(0,-40){2}
\Photon(0,40)(0,-40){8}{4.5}
\Text(0,-45)[]{(a)}
\end{picture}
\hspace{-0.2cm}
\begin{picture}(90,90)(-45,-45)
\SetScale{0.5}                 
\ArrowLine(-80,80)(0,40)
\ArrowLine(0,40)(80,80)
\ArrowLine(0,-40)(-80,-80)
\ArrowLine(80,-80)(0,-40)
\Gluon(-60,70)(60,70){8}{5}
\Vertex(-60,70){2}
\Vertex(60,70){2}
\Vertex(0,40){2}
\Vertex(0,-40){2}
\Photon(0,40)(0,-40){8}{4.5}
\Text(0,-45)[]{(b)}
\end{picture}
\hspace{-0.2cm}
\begin{picture}(90,90)(-45,-45)
\SetScale{0.5}                 
\ArrowLine(-80,80)(0,40)
\ArrowLine(0,40)(80,80)
\ArrowLine(0,-40)(-80,-80)
\ArrowLine(80,-80)(0,-40)
\Vertex(0,40){2}
\Vertex(0,-40){2}
\Photon(0,40)(0,-40){8}{4.5}
\Gluon(-60,70)(-60,-70){-10}{6}
\Vertex(-60,70){2}
\Vertex(-60,-70){2}
\Text(0,-45)[]{(c)}
\end{picture}
\hspace{-0.2cm}
\begin{picture}(90,90)(-45,-45)
\SetScale{0.5}                 
\ArrowLine(-80,80)(0,70)
\ArrowLine(0,70)(80,80)
\ArrowLine(0,-70)(-80,-80)
\ArrowLine(80,-80)(0,-70)
\Vertex(0,70){2}
\Vertex(0,-70){2}
\Photon(0,70)(0,-70){8}{8}
\Gluon(-60,77.5)(60,-77.5){8}{12}
\Vertex(-60,77.5){2}
\Vertex(60,-77.5){2}
\Text(0,-45)[]{(d)}
\end{picture}
\caption{The type of diagrams contributing to the matrix element
of $X$-boson exchange. The wiggly line is the $X$-boson,
the curly line is a gluon and full lines are
quarks.
(a) Wave Function Renormalization.
(b) Vertex Diagrams. (c)(d) Box Diagrams.}
\label{figXdiags}
}
\ba
\label{cutoff}
 i  C_{C}
&& \left[ \left( 1 + F(q_1,q_2,q_3,q_4)
\frac{\alpha_s(\mu_C)}{\pi}  \right) S_1 
+ \left( 1 +  F(q_1,q_2,-q_4,-q_3)
\frac{\alpha_s(\mu_C)}{\pi}  \right) S_2  
\right]\nonumber\\
&&+{\cal O}\left(\frac{1}{M_X^4}\right)
\ea
 using the diagrams of Figure \ref{figXdiags}, with
\be
\label{CC}
C_C
\equiv  \frac{- g_{\Delta S=2}^2}{M_X^2}
 \left(  1 + \frac{\alpha_s(\mu_C)}{\pi} 
\left[ \frac{ \gamma_1}{2} \ln 
\left(\frac{2 q_1 . q_2}{M_X^2}\right)
+ \tilde r_1  \right ] \right) \;.
\ee
We have used the 
conditions in the $X$-boson effective theory $M_X^2 \gg q_i^2$
and $\mu_C^2\gg -q_i^2$.
For the scheme dependent constant we get
\be
\tilde r_1 = -\frac{7}{8} \left(1-\frac{1}{N_c}\right) \, . 
\ee
The box diagrams are finite in this case, the only divergent ones
are the vertex and wave-function renormalization diagrams,
(a) and (b) in Figure \ref{figXdiags}.

The expression in (\ref{cutoff})
is also scale- and scheme-independent 
at order $\alpha_s^2$. The constant
$\tilde r_1$ is scheme\footnote{Here scheme means
the precise definition of the $X$-boson couplings and QCD regularization
used.} dependent but it
cancels against the same
dependence in the constant  $g_{\Delta S=2}$.

About the QCD regularization used, 
notice that there are no logarithms depending on the scale
$\mu_C$. This is not trivial, in principle
these can also appear. This is a consequence of the fact that 
the anomalous dimension vanishes for
two-quark vector and axial-vector currents 
and that box diagrams are finite. 
Two-quark currents can be identified always in the low-energy
approximation to QCD unambiguously as said before.
And for box diagrams we can use $D=4$ 
 and therefore their contribution is $\gamma_5$ and evanescent
operator schemes independent.

The infrared regulator scheme dependence is , as it should be,
precisely the same in Eq. (\ref{DIM}) and (\ref{cutoff}),
precisely the same function $F(q_1,q_2,q_3,q_4)$ appears.

The $X$ boson mass acts here like an ultraviolet regulator.
The dependence has to disappear in the final
physical amplitude. We will see how this happens in the
matching between long- and short-distances.

So from the matching condition in (\ref{match})
 we obtain for the $X$-boson couplings:
\ba
\label{Xcoupling}
g^2_{\Delta S=2}( \mu_C, M_X,\cdots)& =& 
M_X^2 C_{\Delta S=2}
 C(\nu)\Bigg(1+\frac{\alpha_s(\nu)}{\pi}
\left[\gamma_1 \ln\frac{M_X}{\nu}+r_1-\tilde r_1\right]
\nonumber\\&+& 
{\cal O}\left(\alpha_s^2(\mu_C),\alpha_s^2(\nu),\alpha_s(\nu)-\alpha_s(\mu_C)
\right)
\Bigg)\,.
\ea
and $\alpha_s(\nu)-\alpha_s(\mu_C)$ is also of order $\alpha_s^2$.
We also plotted $-g^2_{\Delta S=2}/(M_X^2 C_{\Delta S=2})$
in Figure \ref{figCD} to show the effect of $\tilde r_1$
for the HV and NDR schemes.
Again we see that the result is stable with $\nu$. We used
$M_X=1$~GeV for definiteness.

With this first matching we have
obtained an expression which is scale and scheme independent to NLLO order.
In addition in the $X$-boson theory the only part that needs regularization
is the $X$-boson--quark vertex itself and we have {\em not} used any
$1/N_c$ argument until now.

\subsection{Long-Distance Short-Distance Matching and $1/N_c$}
\label{LDSDtransition}

With the value of $g_{\Delta S=2}$
set in (\ref{Xcoupling})
we can  calculate the weak matrix element
in the $1/N_c$ expansion within the $X$-boson
exchange effective theory.

We follow the same technique we used for
calculating the electromagnetic mass difference
for pions and kaons in \cite{BPem} and related work can be found  
in \cite{BBGpi,Dashen,KPR98}.
We obtained very nice matching for four-point functions in the presence
of quark masses. At present, 
this non-trivial matching has only been obtained in the 
$1/N_c$-technique.

A point we do not discuss here is the choice of gauge for the $X$-boson.
For reasonable gauges the effect are suppressed by extra powers of $1/M_X^2$
so they are not needed here. In any case
the discussion for the photon in \cite{BPem} can be easily extended to show
that there is no problem here in general either.

We want to calculate some matrix element between hadronic states
in general. For our example in (\ref{KKB}):
\ba
\lefteqn{\langle\overline{K^0}|K^0\rangle
= \langle\overline{K^0}|e^{i\Gamma_{\Delta S=2}}|K^0\rangle =
\langle\overline{K^0}|e^{i\Gamma_X}|K^0\rangle}
&&\nonumber\\&=& -  \frac{g_{\Delta S=2}^2}{2}   \langle\overline{K^0}|
\int {\rm d}^4 x \,  \int {\rm d}^4 y \, 4 
L^\mu (x) L_\mu(y) P_X(x,y)|K^0>\,.
\ea
The equality follows from the discussion in the previous subsection. $P_X(x,y)$
is the $X$-boson propagator
\begin{equation}
\label{propX}
P_X(x,y) = \int\frac{d^4p_X}{(2\pi)^4}\frac{i}{p_X^2-M_X^2}
e^{-i(x-y)\cdot p_X}\;.
\end{equation}
The matrix element corresponds to  a four-quark operator
in an OPE in the $X$-boson effective theory.

We now rotate the integral in (\ref{propX}) into the Euclidean
and split the integral into two parts $|p_E|>\mu$ and $|p_E|<\mu$.
One of the reasons to rotate to Euclidean space is that in the lower
part then all components of $p_E$ are also small.
Also, in general, amplitudes are much smoother in the Euclidean region,
so approximations are generally better behaved since threshold effects
and the like are smeared out.
In particular we split
\begin{equation}
\int {\rm d}^4p_E = \int d\Omega_{p_E}\; \Bigg(
\underbrace{\int_0^\mu {\rm d} |p_E| \, |p_E|^3}_{\mbox{long-distance}}
+\underbrace{\int_\mu^\infty {\rm d}|p_E| \,  
|p_E|^3}_{\mbox{short-distance}} \Bigg)\;.
\end{equation}
This is the same procedure as applied to the electromagnetic
mass difference of pions and kaons in \cite{BPem}.
We choose $M_X \gg \mu$ such that in the long-distance part
we can neglect all momentum dependence of the $X$-propagator
and we expand the short-distance part in $1/M_X^2$.

Form-factors of mesons at large Euclidean
momenta $p_E^2$ are suppressed
by $1/p_E^2$, so in the short-distance part the high momentum has to flow
back through quarks and gluons to leading order in $1/\mu^2$.
The short-distance part is a two-step calculation, we evaluate the
the integral with the momentum flowing back through gluons and quarks
using the box diagrams of Figure \ref{figXdiags} only. The vertex diagrams
do not involve a large $X$-boson momentum so they are part of
the long-distance calculation. Here there is no infrared ambiguity,
the possible infrared divergence is regulated by $\mu$, the lower limit
of the short-distance integral.
The result is:
\begin{equation}
\langle\overline{K^0}|{K^0}\rangle_{SD\mbox{-}X}
= \frac{-i g_{\Delta S=2}^2}{2 M_X^2}
\langle\overline{K^0}|S_1+S_2|K^0\rangle
\frac{\alpha_s(\mu)}{\pi} \gamma_1 \ln \frac{\mu}{M_X}\,.
\end{equation}
Here $\alpha_s$ is already suppressed by $1/N_c$ so it is sufficient
to calculate $\langle\overline{K^0}|S_1+S_2|K^0\rangle$
to leading order in $1/N_c$; i.e. $N_c\to \infty$.

Now the long-distance part can be calculated also order by order
in $1/N_c$. The leading order in $1/N_c$ is 
\begin{equation}
\langle\overline{K^0}|{K^0}\rangle_{LD\mbox{-}X-\mbox{leading}}
= \frac{-i g_{\Delta S=2}^2}{2 M_X^2}
\langle\overline{K^0}|S_1+S_2|K^0\rangle_{N_c\to \infty}
\end{equation}
where the subscript $N_c\to \infty$ 
indicates  again the leading in $1/N_c$ contribution.
For the subleading in $1/N_c$ part it is sufficient to replace
$1/(p^2-M_X^2)$ by $-1/M_X^2$ so we obtain
\ba
\label{almostfinal}
\langle\overline{K^0}|{K^0}\rangle & =&
\frac{-ig_{\Delta S=2}^2}{2 M_X^2}
\langle\overline{K^0}|S_1+S_2|K^0\rangle_{N_c\to \infty}
\nonumber  \\
&\times & \left( \frac{\langle\overline K^0|
\int {\rm d}^4 x\int {\rm d}^4 y \, 4 L^\mu(x)
L_\mu(y) P_E(x,y)|K^0>^\mu_{1/N_c\;sup}}
{\langle\overline{K^0}|S_1+S_2|K^0\rangle_{N_c\to \infty}}
+ \frac{\alpha_s(\mu)}{\pi}
\gamma_1 \ln \frac{\mu}{M_X} \right)\,. \nonumber \\
\ea
The subscript $1/N_c\; sup$ means only the $1/N_c$ suppressed part
and 
\begin{equation}
P_E(x,y) = \int_0^\mu {\rm d} |p_E| \, |p_E|^3 
\int {\rm d} \Omega_{p_E} \, e^{-i(x-y)\cdot p_E}.
\end{equation}

We can now insert the value of $g_{\Delta S=2}^2$ 
of Eq. (\ref{Xcoupling}) 
into (\ref{almostfinal}) and we see that all $M_X$ dependence disappears.
The final result is
\begin{eqnarray}
\label{final}
\langle\overline{K^0}|{K^0}\rangle & =&
-\frac{i}{2} C_{\Delta S=2} \, C(\nu)\left(
1+\frac{\alpha_s(\nu)}{\pi}\left[\gamma_1\ln \left(\frac{\mu}{\nu}\right)
+r_1-\tilde r_1\right]\right)
\Bigg(\langle\overline{K^0}|S_1+S_2|K^0\rangle_{N_c\to \infty}
\nonumber\\
&+&\langle\overline{K^0}|\int {\rm d}^4 x \int {\rm d}^4 y 
 \, 4 L^\mu (x) L^\mu(y) P_E(x,y)|K^0>^\mu_{1/N_c\;sup}\Bigg)
\end{eqnarray}

The $\nu$ and $\mu_C$ dependence had already disappeared in the
previous section. The $\mu$ dependence that remains in the overall
coefficient and in the upper limit of the integral in $P_E(x,y)$ are
both at the same order in $1/N_c$ and will cancel if the low-energy
approximation to QCD used to evaluate the long-distance $1/N_c$ suppressed
contribution matches onto QCD sufficiently well.
The complete short-distance corrections to two-loops
can be seen in Figure \ref{figCD}.

Notice that this proof only required the use of the $1/N_c$-expansion
at a rather late stage, thus the arguments all go through provided
the scales $\mu$, $\mu_C$, and $\nu$ are chosen such that no large
logarithms of their ratios appear.
The identification of the short-distance perturbative QCD
quantities with the quantities
in the long-distance $1/N_c$ suppressed part need only be done
at the level of two-quark currents. Once this is done, and as mentioned
before this is normally a requirement for them, the scheme dependence
introduced in short-distances is fully eliminated.

The {\em only}, but difficult,  obstacle for an almost perfect matching
is now to find a model that is viable up to scales
$\mu$ where perturbative QCD is applicable. 

The matching between long- and short-distance
can be systematically improved with low-energy realizations
of QCD which are better and better at high energies.

\subsection{Gluon Gauge Dependence} 
\label{gauge}

In the above we have always used the Feynman gauge. So what happens
now in other gauges. We can investigate the gauge-dependence of both
matchings done before.

The long-distance--short-distance matching in Section \ref{LDSDtransition}
can be easily shown to be gauge-independent to the order we are working.

The transition to the $X$-boson scheme from the operator product expansion
is in fact dependent on the gauge. Actual calculations show that
the gauge dependence is purely a long-distance phenomenon\footnote{See also
the short discussion in the second reference in \cite{BUR98},
Section III.C.}.
The problem is that the state we have chosen in Eq. (\ref{match})
is not a physical
state. We could have chosen an infrared well-defined
observable to do the matching from the OPE to the $X$-boson theory.
This would mean going to on-shell massless quarks and including
soft-gluon radiation as well. Then the gauge independence would have
been explicit.

The other alternative is to use
a gauge-independent infrared regulator but with it fixed at the same
value for both sides or simply, as we did, to use the same gauge
and infrared regulator on both sides of Eq. (3.6).
We have checked that the gauge dependence with our off-shell quarks
is the same on both sides and it does cancel in the value of
$g_{\Delta S=2}$.

\section{Scheme Independent Results for the $B_K$ Parameter}

The kaon bag-parameter $\hat B_K$ is defined by
\begin{equation}
\langle\overline{K^0}|K^0\rangle \equiv
-i C_{\Delta S=2}
\frac{16}{3} f_K^2 m_K^2 \hat B_K\,. 
\end{equation}
The fully scheme-independent result from (\ref{final}) gives
\ba
\label{BKscale}
\hat B_K &=& \, C(\nu) \, 
\left( 1 + \frac{\alpha_s(\nu)}{\pi} 
\left[ \gamma_1  \ln\left(\frac{\mu}{\nu}\right) + r_1 - \tilde r_1 
\right] \right) B_K(\mu)\; ;
\ea
where
\ba
 B_K(\mu)&= & 
 \frac{3}{4}\left( 1 +  \frac{\langle\overline{K^0}|\int {\rm d}^4 x
\int {\rm d}^4 y \, 4 L^\mu (x) L^\mu(y) P_E(x,y)|K^0>^\mu_{1/N_c\;sup}}
{\langle\overline{K^0}|S_1+S_2|K^0\rangle_{N_c\to \infty}} \right)
\ea
with
\be
r_1^{NDR}-\tilde r_1 = -\frac{11}{8} \left(1-\frac{1}{N_c}\right) \, 
\quad\mbox{ and }\quad
r_1^{HV}-\tilde r_1 = -\frac{7}{8} \left(1-\frac{1}{N_c}\right) \, . 
\ee
In our previous work \cite{BP99} we used, as mentioned there,
$\tilde r_1=0$ and $r_1^{NDR}=-7/6$\cite{NLODeltaS=1}. Here we adjust for that.

The remaining matrix element can now be calculated in various ways.
If we use NLO Chiral Perturbation Theory we obtain
in the chiral limit\footnote{The diagrams contributing are those
of Figure 3 in \cite{BP99} but the vertices can now also be those
{}from the order $p^4$ CHPT Lagrangian.}
\begin{equation}
\label{BKCHPT}
B_K^\chi(\mu)=\frac{3}{4}
\left(1-\frac{3\mu^2}{16\pi^2 F_0^2}+\frac{6\mu^4}{16\pi^2 F_0^4}
\left(2L_1+5L_2+L_3+L_9\right)+{\cal O}(\mu^6) \right)\,.
\end{equation}
Away from the chiral limit the expression are much more cumbersome
but for the lowest order CHPT they can be found in \cite{BP99}.
This way we can obtain it as a series in $\mu^2$. But as plotted
in Figure \ref{figBKLD} we see that this series is already breaking
down at values of $\mu\approx 500$ MeV.
We therefore need an extension of CHPT that includes the above result
as a low energy limit. The ENJL model has the same chiral structure
as QCD and also includes spontaneous chiral symmetry breaking.
It thus has CHPT as an automatic built in limit. We then choose
the three ENJL parameters such that the CHPT parameters to order $p^4$
are well described so Eq. (\ref{BKCHPT}) is included. This version
of the model also correctly reproduces a lot more hadronic
phenomenology\cite{ENJL}. In Section \ref{enjl2} we summarize some
of its advantages and disadvantages and how we expect the latter to be
of little influence for the numerical result.
The results for $B_K^\chi(\mu)$, the chiral limit result,
 and $B_K(\mu)$, the result at the quark masses corresponding to
the physical pion and kaon masses are given in Table 1 of \cite{BP99}.
We have plotted these as well in Figure \ref{figBKLD}.
We can see here very well that the ENJL model includes the CHPT results
at low $\mu$ and considerably improves on it at high $\mu$, both in 
the chiral limit and for physical values of the quark masses.
All curves are in the nonet limit.
\FIGURE{
\epsfig{file=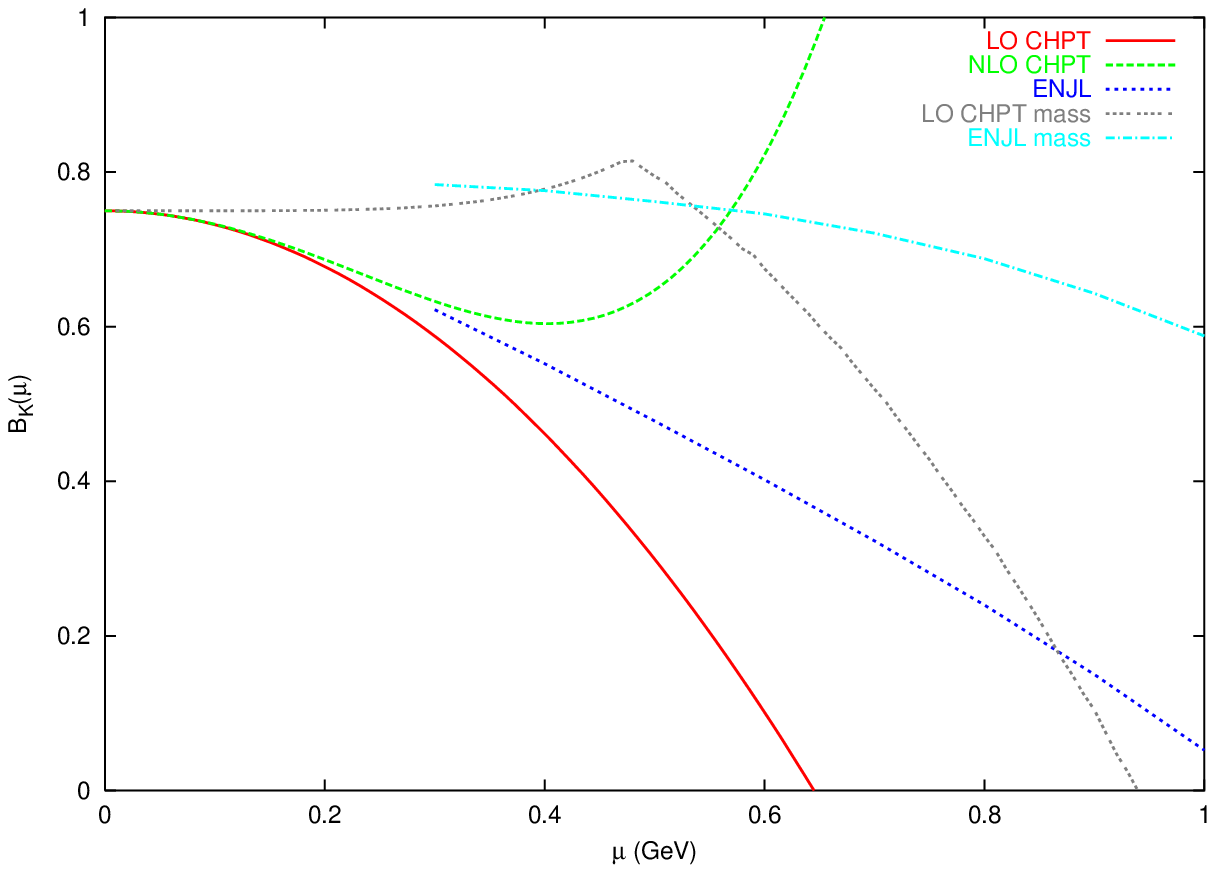,width=\textwidth}
\caption{The long-distance part of the $X$-boson exchange calculation.
We show $B_K(\mu)$ of Eq. (\ref{BKscale}). The lines are leading- (LO) and
next-to-leading-order (NLO) CHPT and the ENJL model for the chiral limit
and LO CHPT and the ENJL model for the physical quark masses.
Notice the improvement of the ENJL model over CHPT.}
\label{figBKLD}
}

We can now put these together with the short distance
part  with  $\alpha_s(M_\tau)=0.334$ and $\mu=\nu$. We use
for definiteness the NDR results for $g_{\Delta S=2}$
but we could have used the HV ones with the same answer, 
see Figure \ref{figCD}.
We also included for the massive case the needed correction
of $0.09\pm0.03$ for the octet versus nonet case\cite{BPBK,BP99},
mainly due to $\eta$-$\eta'$ mixing.
The result is shown as a function of $\nu$ in Figure \ref{figBK1}.
Notice the quality of the matching. 

In \cite{Dortmund} a better identification of the scale 
together with the Leading-Order CHPT approximation 
was used, this  corresponds to keeping the quadratic divergence only.
In the chiral limit this correspond to the first two terms
in (\ref{BKCHPT}). Negative values for $B_K$ and the related
27-plet coupling when using this approximation
in the chiral limit (see Figure \ref{figBK1}), were also seen
there\cite{Dortmund}.
Notice in the same figure how much the use of higher orders estimated 
via the ENJL model
improves the matching in the chiral limit and outside the chiral limit
\cite{BP99}. 
The variation is more due to the $\mu$ dependence since
as shown in Figure \ref{figCD} there is little dependence on $\nu$.
The massive case has an extremely good stability for 700~MeV to 1~GeV
while the chiral case is stable for 550 to 700 MeV. 
\FIGURE{
\epsfig{file=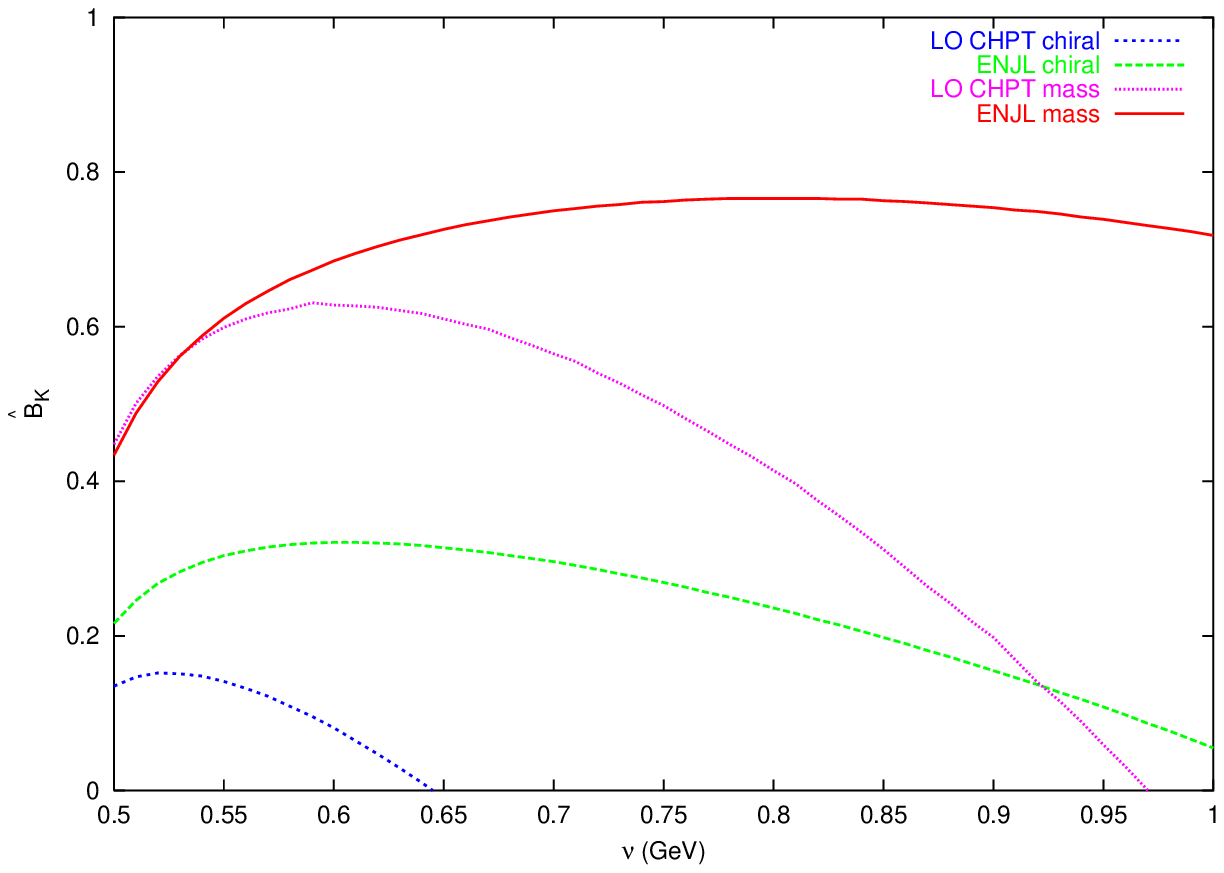,width=\textwidth}
\caption{The results for $\hat B_K$ in the NDR scheme with
 $\alpha_s(M_\tau)=0.334$ as a function of $\mu=\nu$.
The curves are from top to bottom:
ENJL massive, leading-order CHPT massive, ENJL chiral limit,
leading order CHPT chiral limit.
Notice the ENJL stability as compared with leading-order CHPT.
Next-to-leading-order CHPT is even more unstable.}
\label{figBK1}
}

As already noted several times before, \cite{BPBK,BP99} and references therein,
the non-zero quark-mass corrections
to $\hat B_K$ are quite sizable.

We can now study the dependence on the variation with the input.
In Figure \ref{figBK2} we have plotted the result with the same input for
the HV scheme with the same value for $\alpha_s(M_\tau)$
and in the NDR scheme for $\alpha_s(M_\tau)=0.36$ and $0.31$, including
the value of \cite{Toni}.
We took again $\mu=\nu$.
The stability is essentially unchanged while the actual values are very
similar.
\FIGURE{
\epsfig{file=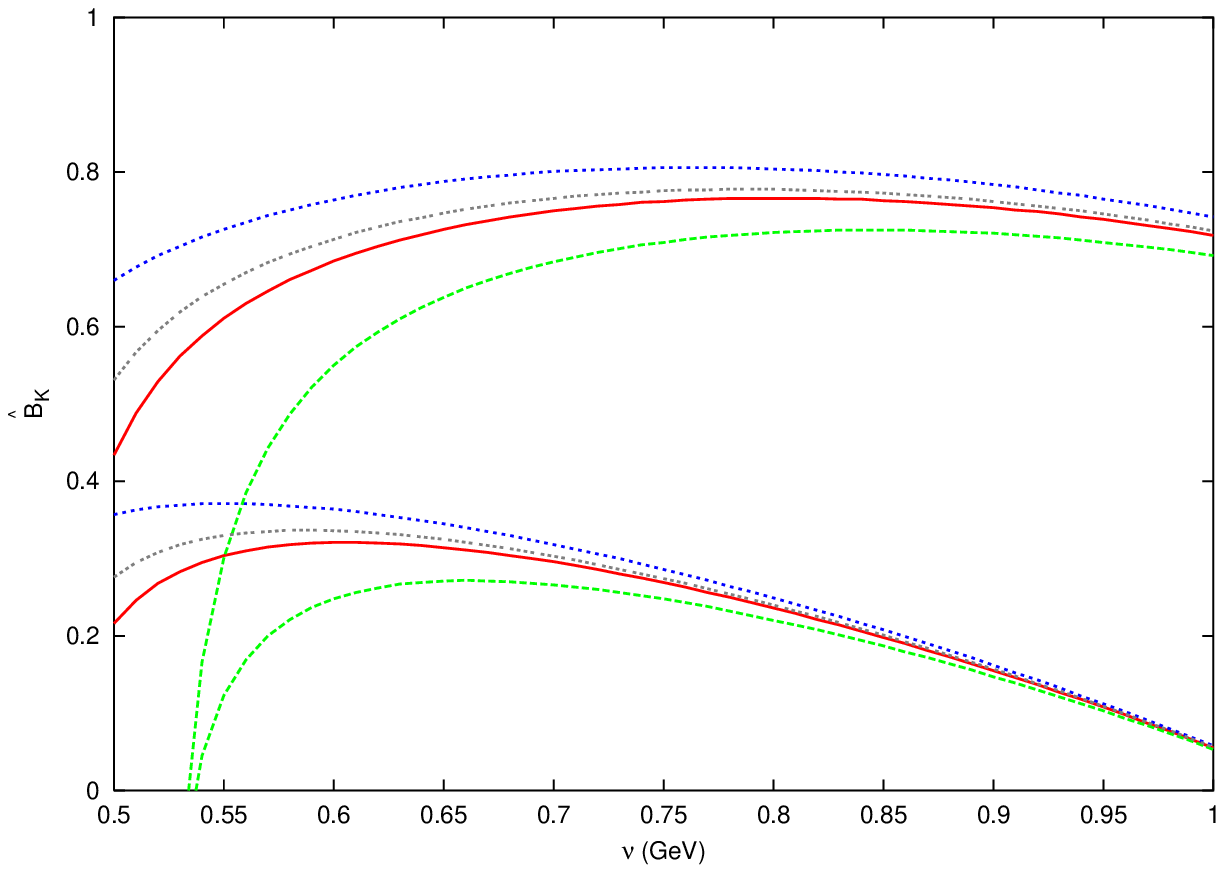,width=\textwidth}
\caption{The variation of $\hat B_K$ with $\alpha_s$
as a function of $\nu$.
The top set of curves is the massive case, the bottom set is
the chiral limit case.
Curves are from top to bottom:
NDR with $ \alpha_s(M_\tau)=0.31$;
HV  with $ \alpha_s(M_\tau)=0.334$;
NDR with $ \alpha_s(M_\tau)=0.334$;
NDR with $ \alpha_s(M_\tau)=0.36$.}
\label{figBK2}
}

Finally we want to check the variation with $\mu\ne\nu$.
This again provides a check on neglected corrections of order $\alpha_s^2$.
In Figure \ref{figBK3} we plotted the NDR case for $\alpha_s(M_\tau)=0.334$
as a function of $\mu$ for, from top to bottom,
$\nu = (1.5,1.2,1,0.8)\mu$. The last curve we did not plot
for low values of $\mu$ since this gets very large values for $\alpha_s(\nu)$.
Again, the variation is rather small.
\FIGURE{
\epsfig{file=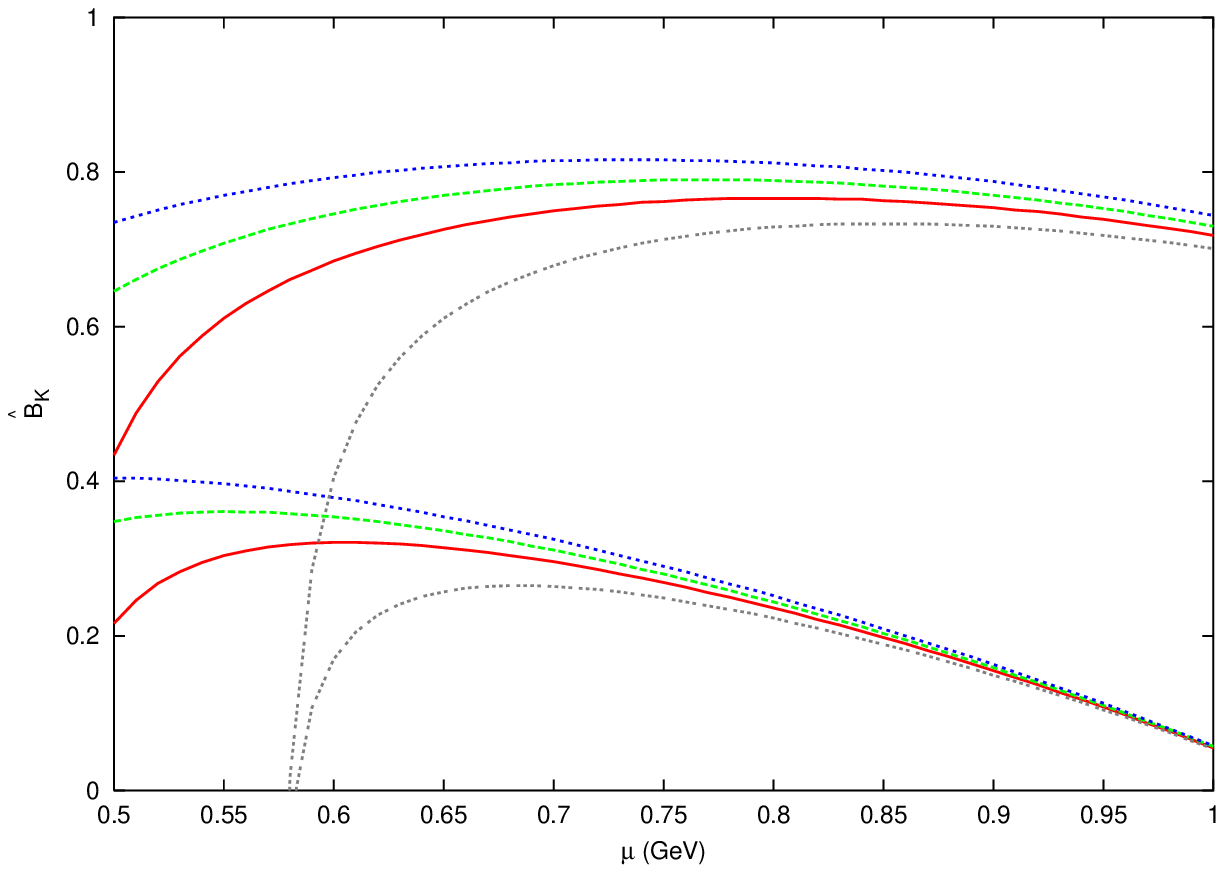,width=\textwidth}
\caption{The variation of $\hat B_K$ in the NDR scheme for, from top to bottom,
$\nu=(1.5,1.2,1,0.8)\mu$. The top set of curves shows the massive case,
the bottom set the chiral limit.}
\label{figBK3}
}

The uncertainty can be judged from all these variations.
We use the variation with $\alpha_s(M_\tau)$ and add a similar
error for the remaining model uncertainty to obtain the
result given in Eq. (\ref{BKresult}).

\section{Some Remarks on the ENJL Model and Possible Improvements}
\label{enjl2}

The model we use  has the chiral structure
of QCD at large $N_c$, i.e. it is a left-right symmetric
model which breaks down to the vector subgroup.
Quark masses break this symmetry as in QCD.

It provides a picture of build-up of constituent quarks out of massless
ones. 
It has two drawbacks, it doesn't confine and in some channels
it does not reproduce the same high-energy behaviour as QCD.
These violations are small at low-energies.
It reproduces CHPT to order $p^4$ \cite{ENJL},
and provides an estimate of all the higher order corrections.
It reproduces a large  
amount of low-energy phenomenology for both the anomalous
and the non-anomalous sectors \cite{ENJL} with an accuracy
at the level of 20\% to 30\%. 

The drawbacks we think will not affect our numerical results too much
because:
\begin{enumerate}
\item We do all our integrations in the Euclidean domain. Here the effect
of sharp states are smeared out so a model without confinement that
reproduces smeared quantities correctly should be all right.
\item We only use the ENJL model at intermediate momenta where CHPT fails
and QCD might not work yet. The good matching we obtain is an indication
that in the regime where we use it the ENJL model works satisfactorily.
\end{enumerate}

The model works all right but is still the weak link in our approach.
Work on  some matrix-elements using more direct QCD based arguments
has been done\cite{KPR98,PPR} and alternatively, we can augment the ENJL model
in ways that improve the matching with QCD by requiring matching
in as many relevant Green functions as possible.

\section{Conclusions}

In this paper we have shown how the short-distance large logarithms can
be resummed in a consistent fashion while retaining the beauty of the
$1/N_c$ approach. We have thus shown how to realize Bardeen's picture
of tracking scheme-dependence all the way \cite{BAR89,BAR99} from
the  $W$-boson mass to very low scales. The approach is general but we
illustrated it in the case of the $B_K$ parameter explicitly.

The main assumption is that we know how to hadronize quark 
currents, once that
is done, our approach tells how the four-quark operators
 should be hadronized. 
This can be systematically improved by higher order calculations within
the perturbative domain.
The low-energy hadronic realization of currents of QCD is the 
only remaining model dependence. 
At very low energies we know that CHPT describes the QCD behaviour
correctly, thus this should be included. In the intermediate domain
we can try to use data as much as possible and/or models with various QCD
constraints. We have argued that the ENJL model is a good first step beyond
CHPT and our final results show this improvement.

By varying $\alpha_s$, the matching scales and conditions, i.e. $\mu\ne\nu$
we can get an estimate for the error involved. Our final result for
$\hat B_K$ is:
\ba
\label{BKresult}
\hat B_K &=& 0.77\pm0.05 \;(\alpha_s)\;\pm0.05\;(\mbox{model});
\nonumber\\
\hat B_K^\chi &=& 0.32\pm0.06 \;(\alpha_s)\;\pm0.12\;(\mbox{model}).
\ea
The difference between these results and the ones in
\cite{BPBK,BP99} is only the short-distance scheme dependence.
Our results here are short-distance scheme independent.
The main uncertainty are the value of $\alpha_s(M_\tau)$,
and a similar error for the remaining
model dependence. The small model dependence error we quote is due
to the almost cancellation of the $1/N_c$ corrections,
remember $\hat B_K=0.75$ at leading order in $1/N_c$.
Possible systematic errors are of course difficult to estimate.
The scale and scheme dependence are consistently
matched at order $\alpha_s^2$ for
current $\times$ current operators and are only a small part of the final
error, especially for the physically relevant massive case.

The scheme dependence has not
been discussed in other non-leptonic weak matrix elements
calculations like the $1/N_c$ technique in 
\cite{Dortmund} or the chiral quark model
in \cite{Trieste}.

 We will apply the same procedure to the
 $\Delta S=1$ transitions where first results
on the $\Delta I=1/2$ rule were obtained in \cite{BP99} using 
 $r_1^{NDR}$ from \cite{NLODeltaS=1} and $\tilde r_1=0$.
Work is in progress to perform the remaining calculations also
for $\Delta S=1$ transitions and
 $\varepsilon'/\varepsilon$\cite{BPP}.

\section*{Acknowledgements}
J.P. would like to thank the hospitality of the Department
of Theoretical Physics at Lund University (Sweden) where part 
of his work was done. This work has been supported in part by 
the European Union TMR Network EURODAPHNE (Contract No. 
ERBFMX-CT98-0169), by CICYT, Spain, under Grant No. AEN-96/1672,
 by Junta de Andaluc\'{\i}a Grant No. FQM-101
and by the Swedish Science Foundation (NFR).

\end{document}